\documentclass[aps,prd,preprint,tightenlines,superscriptaddress,
amsfonts,amssymb,amsmath,showpacs,nofootinbib]{revtex4-1}

\usepackage[polish,english]{babel}
\usepackage[T1]{fontenc}
\usepackage[cp1250]{inputenc}
\usepackage{latexsym}
\usepackage{graphicx}
\usepackage{epstopdf}

\newcommand{\bi}{\begin{itemize}}
\newcommand{\ei}{\end{itemize}}
\newcommand{\p}{\partial}

\DeclareMathOperator{\sign}{sign}
\usepackage{xcolor}
\usepackage{pdfsync}
\usepackage{fancyhdr}
\usepackage{makeidx}
\usepackage[breaklinks,colorlinks,backref]{hyperref}
\usepackage[normalem]{ulem}
\hypersetup{
    colorlinks, 
    linktoc=all, 
    linkcolor=red,  
  citecolor=hyptxt,
  urlcolor=blue}
  \hypersetup{
  citebordercolor=,
  filebordercolor=green,
  linkbordercolor=blue
}
\definecolor{hyptxt}{rgb}{0.7, 0.4, 0.9}

\newcommand{\dR}{\mathbb R}

\newcommand{\be}{\begin{equation}}
\newcommand{\ee}{\end{equation}}

\newcommand{\ket}[1]{|\kern.3ex#1\kern.3ex\rangle}
\newcommand{\bra}[1]{\langle\kern.3ex #1 \kern.3ex|}
\newcommand{\scalar}[2]{\langle\kern.3ex #1 \kern.3ex|\kern.3ex#2\kern.3ex\rangle}
\newcommand{\norm}[1]{\|\kern.3ex#1\kern.3ex \|}
\def\be{\begin{equation}}
\def\ee{\end{equation}}
\def\bn{\begin{enumerate}}
\def\en{\end{enumerate}}
\newcommand{\nn}{\nonumber}
\newcommand{\bea}{\begin{eqnarray}}
\newcommand{\eea}{\end{eqnarray}}
\newcommand{\ba}{\begin{array}}
\newcommand{\ea}{\end{array}}
\newcommand{\bl}{\begin{align}}
\newcommand{\el}{\end{align}}
\newcommand{\iind}{$2^\mathrm{nd}~$}

\newcommand{\ivth}{$4^\mathrm{th}~$}
\newcommand{\al}{\alpha}

\newcommand{\om}{\omega}
\newcommand{\g}{\gamma}

\newcommand{\ph}{\varphi}
\renewcommand{\th}{\theta}
\newcommand{\ep}{\varepsilon}
\newcommand{\bs}{\begin{subequations}}
\newcommand{\es}{\end{subequations} \noindent}
\newcommand{\half}{\tfrac{1}{2}}

\renewcommand{\bn}{\begin{enumerate}}
\renewcommand{\en}{\end{enumerate}}

\renewcommand{\textsc}{\sout}
\renewcommand{\textsl}{\textcolor{magenta}}
\renewcommand{\emph}{\textcolor{green}}

\def\half{\frac{1}{2}}

\newcommand{\Komentarz}[1]{}  

\begin{document}


\title{Generic instability of the dynamics underlying \\the Belinski-Khalatnikov-Lifshitz scenario}

\author{Piotr Goldstein} \email{piotr.goldstein@ncbj.gov.pl}
\affiliation{Department of Fundamental Research, National Centre for Nuclear
  Research, Pasteura 7, 02-093 Warsaw, Poland}

\author{W{\l}odzimierz Piechocki} \email{wlodzimierz.piechocki@ncbj.gov.pl}
\affiliation{Department of Fundamental Research, National Centre for Nuclear
  Research, Pasteura 7, 02-093 Warsaw, Poland}

\date{\today}

\begin{abstract}

A class of exact solutions  to the
Belinski-Khalatnikov-Lifshitz (BKL) scenario is derived and tested
for their stability against small perturbations. These are the only
regular solutions in the Painlev\'{e} sense. We prove that they are
unstable in the vicinity of the cosmological singularity. Regularity
of the dynamics is also examined with the dynamical systems method.
Our results confirm the conjecture of BKL that the dynamics near the
singularity becomes generically chaotic.
\end{abstract}



\maketitle

\tableofcontents

\section{Introduction}

By the BKL scenario, we mean  the scenario proposed by Belinski,
Khalatnikov, and Lifshitz to describe the evolution of the
universe towards the cosmological singularity. This scenario,
derived within the general relativity, leads to the conclusion
that the Einstein equations imply existence of a generic solution
with gravitational singularity \cite{BKL2,BKL3}. By the generic
solution the authors mean, roughly speaking, that it
corresponds to a non-zero measure subset of all initial data and
depends on the proper number of arbitrary functions of space.

The derivation of this scenario is based on the general
(non-diagonal) Bianchi VIII and IX models of spacetime
evolving towards the singularity. That dynamics can be simplified
by assuming that some
stress-energy tensor components may be ignored, some Ricci tensor
components have negligible influence, and
anisotropy of space may grow without bounds. These assumptions
lead to enormous simplification of the mathematical form of the
dynamics. It can be well approximated by the following system of
equations \cite{bkl,Belinski:2014kba,book}:
\begin{equation}\label{L1}
\frac{d^2 \ln a  }{d t^2} = \frac{b}{a}- a^2,~~~~\frac{d^2 \ln b
}{d t^2} = a^2 - \frac{b}{a} + \frac{c}{b},~~~~\frac{d^2 \ln c }{d
t^2} = a^2 - \frac{c}{b},
\end{equation}
subject to the constraint
\begin{equation}\label{L2}
\frac{d\ln a}{dt}\;\frac{d\ln b}{dt} + \frac{d\ln
a}{dt}\;\frac{d\ln c}{dt} + \frac{d\ln b}{dt}\;\frac{d\ln c}{dt} =
a^2 + \frac{b}{a} + \frac{c}{b} \, ,
\ee
where $\,a=a(t),\, b=b(t)$ and $\,c=c(t)$ are the so-called
directional scale factors, while $t$ is the time parameter
in the synchronous reference system. Their evolution defines the
dynamics of the characteristic lengths in three directions while
the universe tends to the singularity. Due to time-reversibility
of equations \eqref{L1}--\eqref{L2}, it may also describe
expansion of the universe away from the singularity.

These scale factors include contributions from standard matter
fields, e.g., the perfect fluid with equation of state $p = k
\varepsilon$, $0\leq k < 1$, where  $p$ and $\varepsilon$ denote,
the pressure and energy density of the fluid, respectively. The
case $k=1$ (describing, e.g.,  massless scalar field) is excluded
as it does not lead to oscillatory dynamics inherent in the
Bianchi models. It is likely that other gravity sources may lead
to the asymptotic form \eqref{L1}--\eqref{L2} as well. For
instance, it may include an electromagnetic field or the
Yang-Mills fields, but further examination is required to confirm
that expectation (see, Sec. \!\!4 of \cite{book} for discussion of
this issue).

The system of equations \eqref{L1}--\eqref{L2} has never been
solved analytically, in spite of its importance in the context of
the BKL scenario. In this article, we find an explicit solution to
this dynamics, analyse its stability, regularity, and provide its
physical interpretation.

The BKL scenario \cite{BKL2,BKL3} proposes a mechanism that
leads to the generic singularity via a stochastic process. We
confirm existence of this scenario by showing that the only
regular solution, in the Painlev\'{e} sense, to the
dynamics \eqref{L1}--\eqref{L2} is unstable and leads to chaos.

The paper is organised as follows: In Section II the exact
solution is presented and its stability is examined. The
regularity analysis of the dynamics is carried out in Sec. III
within the dynamical systems method.  The last section presents
the conclusions. The Appendices contain one of the possible
derivations of our solution (App. A) and discuss the issue of the
monotonicity of the space volume (App. B).

\section{Solution}

\subsection{Special exact solution}

The analytical solution of Eqs. \!\eqref{L1}--\eqref{L2} reads
\be\label{solution}
a(t)= \frac{3}{\lvert t-t_0\rvert },~~ b(t)= \frac{30}{
\lvert t-t_0\rvert ^{3}},~~ c (t)= \frac{120}{\lvert t-t_0\rvert
^{5}} \, ,
\ee
where   $\lvert t - t_0\rvert  \neq 0$ and $t_0$ is an
arbitrary real number.

This solution may be obtained by a systematic method rather than a
smart guess. For instance, one can use: (i) extension of the
Painlev\'e test applied to equations \eqref{L1}--\eqref{L2}, (ii)
expansion of these equations about $t=\infty$, or (iii)  a search
for their self-similar solution. In Appendix A, we describe the
first of these methods.

Eqs. \!\eqref{L1}--\eqref{L2} have been derived from the general
dynamics of the Bianchi VIII and IX models under the condition
that near the singularity one has \cite{bkl}):
\be\label{order}
a\gg b\gg c > 0 \, .
\ee
Therefore, the physically relevant part of the special solution to
\eqref{L1}--\eqref{L2} should satisfy that condition as well. Our
solution \eqref{solution} satisfies this condition, provided that
$\lvert t-t_0\rvert$ is sufficiently large, which is true near the singularity
(corresponding to $\lvert t\rvert\to\infty$).

\subsection{Canonical structure}

It is shown in \cite{PC} that equations \eqref{L1} can be derived
from the Lagrangian
\be
\mathcal{L}=\dot{x}_1\dot{x}_2+\dot{x}_2\dot{x}_3+\dot{x}_3\dot{x}_1+\exp(2x_1)+\exp(x_2-x_1)+\exp(x_3-x_2),
\ee
where the dot over a symbol denotes its time derivative $d/dt$, and
\be\label{x}
x_1=\ln a,\quad x_2=\ln b,\quad x_3=\ln c.
\ee

 The dynamical constraint \eqref{L2} corresponds to the
condition that the ``energy''
\be
\mathcal{H}=\sum_{i=1}^3\frac{\p\mathcal{L}}{\p
\dot{x}_i}\dot{x}_i-
\mathcal{L}=\dot{x}_1\dot{x}_2+\dot{x}_2\dot{x}_3+\dot{x}_3\dot{x}_1-\exp(2x_1)-\exp(x_2-x_1)-\exp(x_3-x_2)
\ee
vanishes \cite{PC}.

In order to interpret our solution in terms of the
canonical variables, we perform an orthogonalisation of the
``kinetic'' part of the energy by a linear transformation
\be\label{ortho}
x_1=u_1-u_2-u_3,\quad x_2=u_1+2u_3,\quad x_3=u_1+u_2-u_3 \, ,
\ee
which yields the Lagrangian in the form
diagonal in the ``velocities''
$\dot{u}_1,\,\dot{u}_2,\,\dot{u}_3$
\be\label{Lagr}
\mathcal{L}=3\dot{u}_1^2-\dot{u}_2^2-3\dot{u}_3^2+\exp\big(2(u_1-u_2-u_3)\big)+\exp(u_2-3u_3)+\exp(u_2+3u_3).
\ee
It also provides an analogous expression for the ``energy''
$\mathcal{H}$, which differs from the Lagrangian $\mathcal{L}$ by
having the opposite signs of the ``potential'' part, which
consists of the terms with exponential functions. Thus the
variables $u_1,\,u_2,\,u_3$ define the principal directions in the
velocity space. If the ``energy'' $\mathcal{H}$ is
expressed in terms of the momenta $p_i=\p \mathcal{L}/\p
\dot{u}_i, i=1,2,3$, then it becomes the Hamiltonian, which is
also diagonal in the momenta.

Since the ``energy'' is zero, the whole dynamics of the system
takes place in the inner part of the cone, which means

\be\label{cone1}
3\dot{u}_1^2-\dot{u}_2^2-3\dot{u}_3^2>0 \, ,
\ee
as shown in Fig. 1.
\begin{figure}\label{ccone1}
\begin{center}
\includegraphics[width=0.7\textwidth]{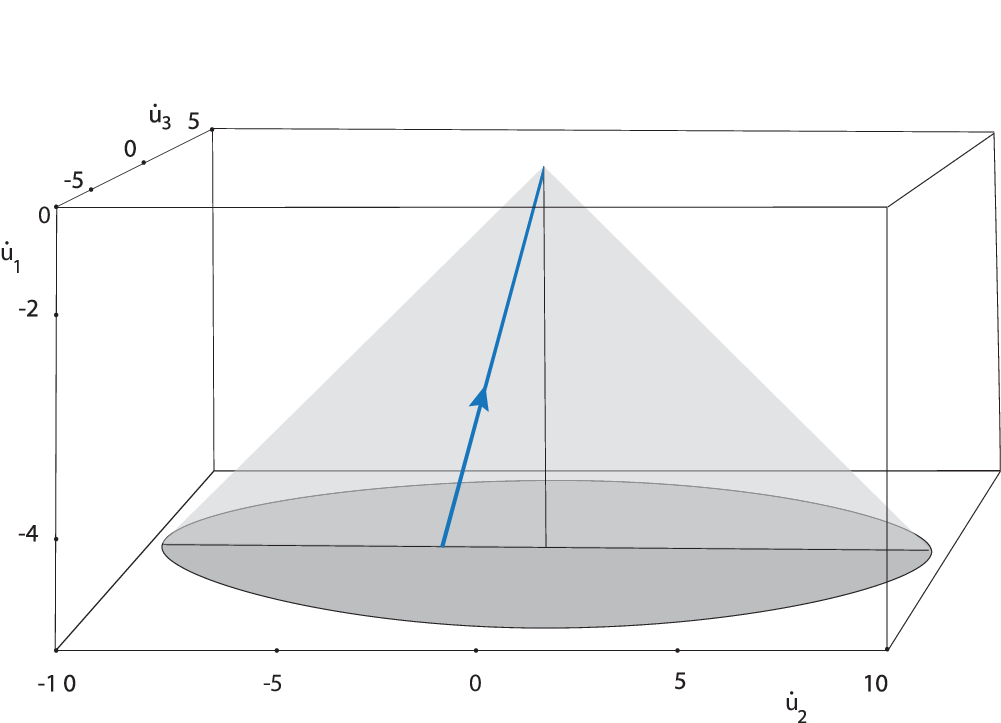}
\caption{The lower half-cone with
$3\dot{u}_1^2-\dot{u}_2^2-3\dot{u}_3^2  > 0$. The dynamics of the
system takes place inside the cone. The exact solution
\eqref{solution} is shown as the line (the arrow indicates the
direction of evolution), which lies in the axial plane
$\dot{u}_3=0$ and satisfies $\dot{u}_2=\tfrac23\dot{u}_1$ . For
$t\to\infty$, the line tends to the vertex of the cone. The other
end of the line extends to infinity (which is beyond the scope of
the BKL scenario).}
\end{center}
\end{figure}

Expression \eqref{cone1} is zero on the conical surface. Together
with the constraint $\mathcal{H}=0$, it means that the exponential
terms in \eqref{Lagr} turn to zero, which is possible only for all
$u$'s tending to $-\infty$. The latter may happen only for
$t\to\infty$ (for the case $t>t_0$).

The exact solution \eqref{solution} corresponds to
\be\label{sol-u}
u_1=\frac13\ln\frac{10800}{\lvert t-t_0\rvert^9} ,\quad u_2=\half
\ln\frac{40}{\lvert t-t_0\rvert^4},\quad u_3=\frac16\ln\frac52,
\ee
Note that the dynamics leaves the third component, $u_3$
unchanged, which means that the evolution is two-dimensional.
Geometrically, the exact solution \eqref{solution} describes the
dynamics within a planar coaxial section $\dot{u}_3=0$ of the cone
\eqref{cone1}, along the line $\dot{u}_2=\frac23\dot{u}_1$, which
lies inside the cone (see Fig. 1).  For $t\to\infty$, the solution
\eqref{sol-u} tends to the vertex of the cone as $\lvert
t-t_0\rvert^{-1}$. The line escapes to infinity, at some finite
$t_0$, also as $\lvert t-t_0\rvert^{-1}$, however this region lies
beyond the range of applicability of the BKL scenario.

Condition $u_3=const.$ of \eqref{sol-u} in terms of the original
variables $a,b,c$ is equivalent to the requirement that $b$ is
proportional to the geometric mean of $a,c$, i.e. $b=k \sqrt{ac}$
with a constant coefficient ($k=\half\sqrt{10}$). This property is
easy to notice in \eqref{solution}. A closer insight shows that
solution \eqref{solution} is the only one satisfying the
geometric-mean condition and the constraint \eqref{L2}.

One can conclude, a posteriori, that  \eqref{solution} is a special
solution to the dynamics \eqref{L1}--\eqref{L2} corresponding  to the
following initial data:
\bl\label{initial}
\nonumber  a(0) &= -3 \;t_0^{-1},&\dot{a} (0) &= -3\; t_0^{-2}\,, \\
b(0) &= -30\;t_0^{-3},& \dot{b}(0) &= -90\; t_0^{-4}\,, \\
\nonumber  c(0) &= -120\; t_0^{-5},& \dot{c}(0) &=  -600\;
t_0^{-6}\,,
\end{align}
for instance, in the case  $t>t_0$ and $t_0 < 0$.


\subsection{Stability analysis}

In what follows, we consider a linear approximation to the general
solution in terms of a small perturbation of the solution
\eqref{solution}.

To check how the small perturbation to the solution
\eqref{solution} develops in time, we substitute the following
functions into \eqref{L1}--\eqref{L2}
\bs\label{pert}
\bl
& a (t) =3(t-t_0)^{-1}+\ep\al(t),\\
& b (t) =30(t-t_0)^{-3}+\ep\beta(t),\\
& c (t)=120(t-t_0)^{-5}+\ep\g(t).
\end{align}
\es
In the first order in the small parameter $\ep$,  we obtain
\bs\label{pert-eqs}
\bl
&\ddot{\al}+\frac{2\dot{\al}}{t-t_0}+\frac{28\al}{(t-t_0)^2}-\beta=0 ,\\
&\ddot{\beta}+\frac{6\dot{\beta}}{t-t_0}+\frac{20\beta}{(t-t_0)^2}-\frac{280\al}{(t-t_0)^4}-\g=0 ,\\
&\ddot{\g}+\frac{10\dot{\g}}{t-t_0}+\frac{24\g}{(t-t_0)^4}-\frac{16\beta}{(t-t_0)^4}-\frac{720\al}{(t-t_0)^6}=0 ,
\end{align}
with the constraint
\bl\label{pert-cons}
\dot{\g}+\frac{6\g}{t-t_0}+\frac{6\dot{\beta}}{(t-t_0)^2}+\frac{24\beta}{(t-t_0)^3}
+\frac{80\dot{\al}}{(t-t_0)^4}+\frac{160\al}{(t-t_0)^5}=0 .
\end{align}
\es
The system of equations \eqref{pert-eqs} is linear and
homogeneous. Its general solution may be simply written in terms
of the rescaled evolution (time) parameter $\th=\ln\lvert
t-t_0\rvert$ and two frequencies
\be\label{freqs}
\om_1 = \half\sqrt{95-24\sqrt{6}},\qquad\om_2 =
\half\sqrt{95+24\sqrt{6}}.
\ee

The solution, with 6 arbitrary constants: $K_1,...,K_4$ and
$\ph_1,\ph_2$, reads
\bs\label{asymp}
\bl
\al =&\exp(-\th/2)\!\left[K_1\cos(\om_1\th\!+\!\ph_1)\! +\!
K_2\cos(\om_2\th\!+\!\ph_2)\right]
\!+\!K_3\exp(-2\th)\!+K_4\exp(\th) ,\\
\beta=&\exp(-5\th
/2)\left[\left(4+6\sqrt{6}\right)K_1\cos(\om_1\th+\ph_1)\right.\nn\\&\left.+\left(4-6\sqrt{6}\right)K_2\cos(\om_2\th+\ph_2)
\right]+30K_3\exp(-4\th)+30K_4\exp(-\th) ,\\
\g=&-4\exp(-9\th
/2)\left[\left(26+9\sqrt{6}\right)K_1\cos(\om_1\th+\ph_1)\right.\nn\\&\left.+\left(26-9\sqrt{6}\right)K_2\cos(\om_2\th+\ph_2)
\right]+200 K_3\exp(-6\th)+200K_4\exp(-3\th) \, .
\end{align}
To comply with equation \eqref{L2}, we have to impose the
constraint \eqref{pert-cons}, which leads to simple
\be K_4=0.
\ee
\es
\begin{figure}\label{inst-abc}
\centering
\includegraphics[width=135mm]{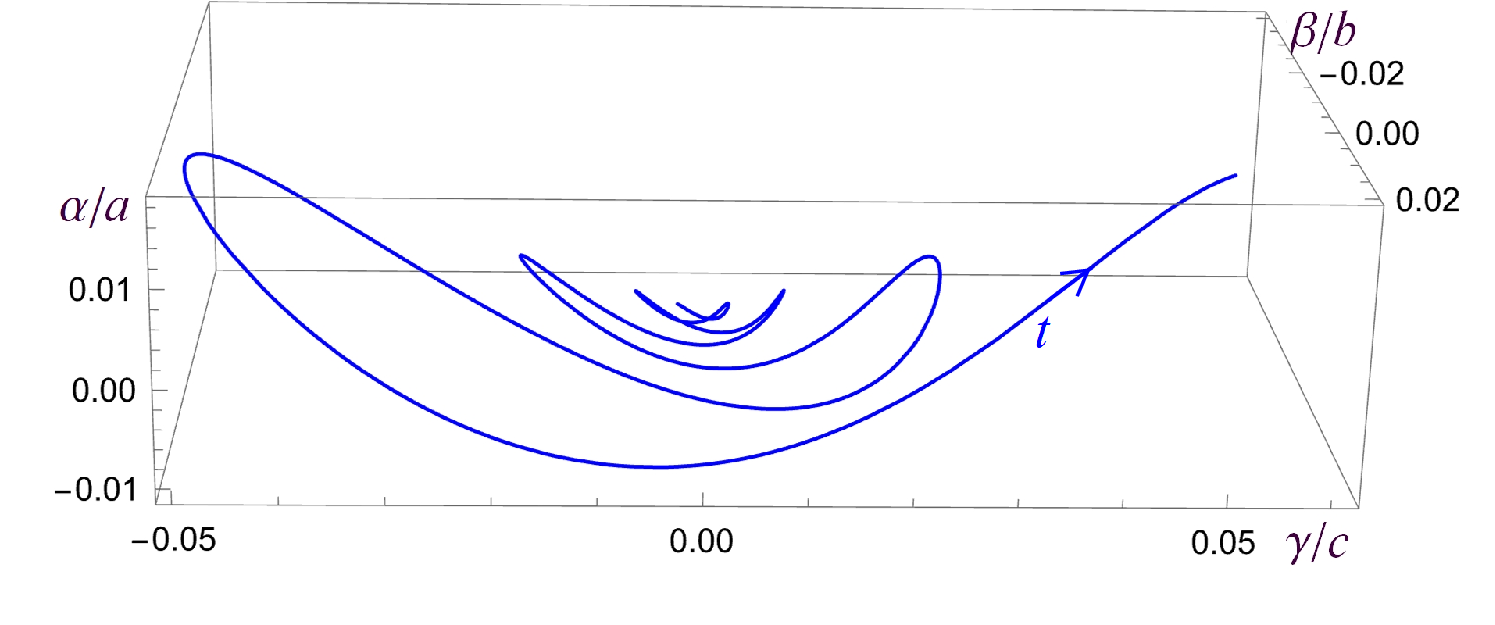}
\caption{Linear instability of the special solution
\eqref{solution} for $K_1=K_2=0.01,\,K_3=0,\,\ph_1=\ph_2=0$ for
the perturbations of the scale factors ($\alpha,\, \beta,\, \g$),
respectively. The graph presents the parametric curve defined by
the time dependence of $\alpha/a,\,
\beta/b$, and $\gamma/c$.}
\end{figure}

Fig. 2 presents the special solution \eqref{solution} in terms of
the time dependence of $\alpha/a,\,\beta/b$, and $\gamma/c$.

 The result \eqref{asymp} shows that for $t\to t_0$
($\th\to -\infty$), the oscillating part of the perturbation to
the solution \eqref{solution} remains small with respect to the
respective components of the solution, although its frequency
increases in the time parameter $t$.  The most important
asymptotic, corresponding to the approach to the singularity
is $t\to\infty$ ($\th\to\infty)$. The fact that
$\al,\,\beta$, and $\g$ tend to zero in this limit does not mean
stability. Solution \eqref{solution} is unstable, in the sense
that the perturbations oscillate with an amplitude greater by a
factor proportional to $(t-t_0)^{1/2}=\exp(\th/2)$ compared to the
perturbed components. In terms of $\th$, the perturbations
oscillate with two characteristic frequencies $\om_1$ and $\om_2$
given by \eqref{freqs}.

The relative perturbations $\alpha/a,\,
\beta/b$, and $\gamma/c$ are absolute perturbations of the
logarithmic variables $x_1,\, x_2$ and $x_3$ defined in \eqref{x}.
We have shown that they grow proportionally to $\exp(\frac12\th)$
in the logarithmic time $\th$. This may be regarded as a
divergence of the neighbouring solutions while the multiplier
$\frac12$ plays the role of a Lyapunov exponent, describing the
rate of their divergence, the same for all three variables $x_1,\,
x_2$ and $x_3$. The positive value of the exponent infers chaotic
evolution of the system (towards the singularity). The transition
to chaos would happen if the evolution started close to our
initial condition \eqref{initial}, within its small neighbourhood
in the 5-dimensional space spanned by the initial values
$a(0),\,b(0),\,c(0)$ and two derivatives (the third one is
uniquely determined by constraint \eqref{L2}).

The incommensurability of the frequencies $\om_1$ and $\om_2$
results in the ergodic character of the evolution which starts
from solution \eqref{solution} perturbed as
in \eqref{pert}. Namely, the phase spaces of $(\al, \dot{\al})$,
$(\beta, \dot{\beta})$, and $(\g, \dot{\g})$ are densely covered
with the trajectories of the perturbed solution for almost all
initial conditions (the exceptions are of measure zero).
Obviously, the 6-dimensional phase space of all three variables
and their time derivatives, even though reduced to five dimensions
by constraint \eqref{L2}, cannot be densely covered by the
two-frequency sinusoidal oscillations.

From the point of view of dynamical systems, this means that
evolution of the instability leads to chaotic behaviour. This way,
even the only regular solution, unique up to translations in $t$,
decays to chaos. This confirms the conjecture of BKL that the
chaotic behaviour inevitably accompanies approach to the
cosmological singularity.

 Eq. \!\eqref{asymp} presents the general solution of
\eqref{pert-eqs}. It depends on three arbitrary real constants $\{
K_1, K_2, K_3 \}$ (expected to be small to  comply with the linear
approximation), and any two real constants $\{\varphi_1, \varphi_2
\}$ from the interval $[0,\, 2\pi[$. The manifold $\mathcal{M}$
defined by $\{ K_1, K_2, K_3, \varphi_1, \varphi_2 \}$ is a
submanifold of $\dR^5$. The solution defined by \eqref{pert} and
\eqref{asymp} corresponds to the choice of the set of the initial
data $\mathcal{N}$ that is a small neighbourhood of the initial
data \eqref{initial}. $\mathcal{N}$ is a submanifold of $\dR^5$ as
\eqref{initial} defines five independent constants due to the
constraint \eqref{L2}. Thus, Eq. \!\eqref{asymp} presents a
generic solution to \eqref{pert-eqs}, in the sense mentioned in
Introduction, as the measures of both $\mathcal{M}$ and
$\mathcal{N}$ are nonzero (although the exact solution
\eqref{solution} is obviously of zero-measure in the space of all
possible solutions of \eqref{L1}--\eqref{L2}).

In this context, the term ``generic'' requires some comments.
Usually, this term is regarded as a more
precise equivalent to ``typical''. However, in the vast see of
solutions to Einstein's equations, there may exist many typical
islands. Therefore, the authors of \cite{BKL2,BKL3} formulated
their conditions of the dependence on sufficient number of
arbitrary functions on space and a nonzero measure of the set of initial
data.

Our ``genericness'' is much more modest as the BKL scenario
consists of ordinary differential equations. Moreover, the
perturbation of each of the scale factors $a,\, b,$ and $c$ is
proportional to the small parameter $\ep$. This makes the nonzero
measure a small quantity vanishing as $\ep\rightarrow 0.$
Hence, our family of solutions is not typical, though it meets the
BKL  criterion of being generic.

For the development of the singularity, an important quantity is
the evolution of the space volume $V = a b c$. The perturbed
volume corresponding to solutions \eqref{solution} and
 \eqref{asymp}, up to first-order terms in the perturbations
$\{\al,\beta,\g\}$, reads \footnote{ We have incorporated the small
parameter $\ep$ into $\al,\,\beta$, and $\g$. }
\begin{align}\label{Vol}
&V = (a+\alpha)(b+\beta)(c+\g) \simeq abc + ab\g + bc\al + ca\beta \nn\\
&= 10800 \exp(-9\th)\left\{1\! +\! \left[Q_1\cos(\om_1
\th\!+\!\ph_1)\!+\!Q_2\cos(\om_2
\th\!+\!\ph_2)\right]\exp(\th/2)\!+\!Q_3\exp(-\th)\right\},
\end{align}
where
\be
Q_1 = -(2/5)\left(4+\sqrt{6}\right)K_1,\quad Q_2 =
-(2/5)\left(4-\sqrt{6}\right)K_2,\quad Q_3=3K_3.
\ee
As seen from \eqref{Vol}, the volume tends to zero for
$\th\to\infty$. However it apparently oscillates with the same two
characteristic frequencies $\om_1$ and $\om_2$ as $a,\, b$ and
$c$. The ratio of the perturbation to the zero order term grows as
$\exp(\th/2)$, as that in the evolution in each direction
\eqref{asymp}.
\begin{figure}\label{inst-V}
\includegraphics[width=0.7\linewidth]{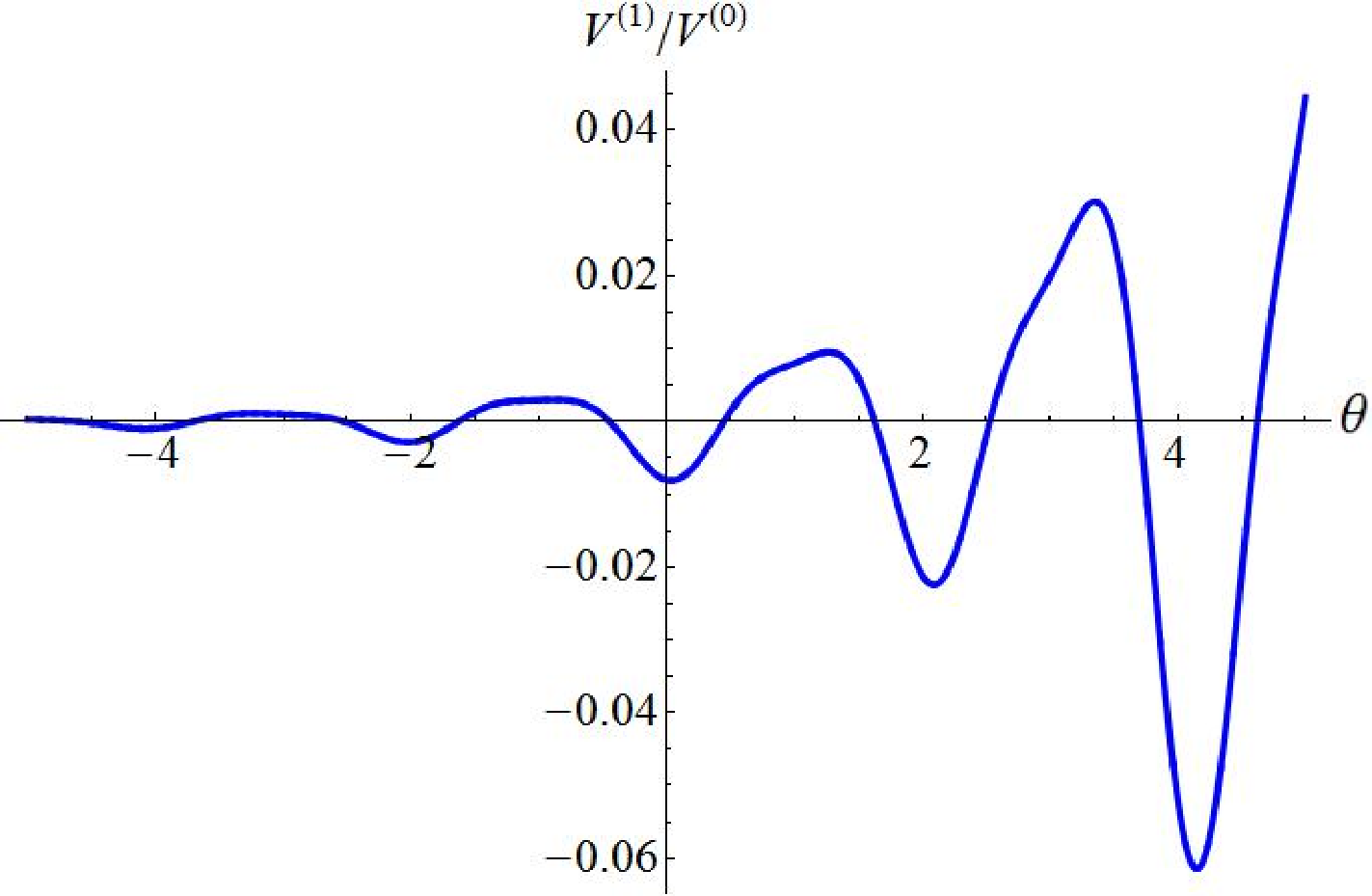}
\caption{Linear instability of the special solution
\eqref{solution} for $K_1=K_2=0.01,\,K_3=0,\,\ph_1=\ph_2=0$. The
graph shows the oscillations of the relative space volume
perturbation $V^{(1)}/V^{(0)}$ as a function of $~\th = \ln\lvert
t-t_0\rvert$, without taking into account the accompanying uniform
expansion. The amplitudes of the two-frequency oscillations
systematically increase as $\exp(\th/2)= \lvert t-t_0\rvert^{1/2}$
in the evolution towards the gravitational singularity
($t\to\infty$). The time interval shown in the graph corresponds
to the range of applicability of the linear approximation,
where $\al,\,\beta,\,\g$ and $V^{(1)}$ remain small fractions of
$a,\,b,\,c$ and $V^{(0)}$, respectively. The nonphysical bouncing
of the volume may be overcome by introducing nonzero, sufficiently
large $K_3$.}
\end{figure}
Fig. 3 presents the instability of the space volume without
showing the accompanying uniform expansion (i.e. for $K_3=0$). The
apparent oscillations are nonphysical because they correspond to
bounces of the space, which contradicts the gravitational
singularity inherent in the BKL scenario. However it turns out
that we can get rid of the bouncing if we restrict the manifold
$\mathcal{N}$ to some submanifold $\mathcal{K}$.
Namely, these volume oscillations
\eqref{Vol} do not affect the actual monotonicity of $V(\th)$,
provided that constant $K_3$ is sufficiently large compared to
$K_1$ and $K_2$. Appendix B presents a proof of this desired
feature of our solution \eqref{asymp}. The evolution of the system
towards the singularity manifests monotonic decrease of the
volume, whereas the evolution away from the singularity is
described by its monotonic increase. Therefore, the linear
perturbation of the volume has the same properties as the
unperturbed volume $V = abc$ corresponding to solution
\eqref{solution}.

\section{Dynamical systems analysis}

In this section we examine the stability of the dynamics by using
the dynamical systems method \cite{Per,Wig}. First, we determine
the critical points of the dynamics (\ref{L1pol})--(\ref{L2pol}),
which characterises the local geometry of the space of solutions.
 For this purpose, we rewrite the system
(\ref{L1pol})--(\ref{L1polc}) in the form suitable for the
analyses:

\begin{align}
  \dot{x} &= x^2 /a + b - a^3 \,, \label{cr1}  \\
  \dot{y} &= y^2 /b + a^2 b - b^2 /a + c \,, \label{cr2}  \\
  \dot{z} &= z^2 /c + a^2 c - c^2 /b \,, \label{cr3} \\
  \dot{a} &= x \,, \label{cr4} \\
  \dot{b} &= y \,, \label{cr5} \\
  \dot{c} &= z \,, \label{cr6}
\end{align}
and the constraint \eqref{L2pol} reads
\begin{equation}\label{cr7}
  cxy + bxz + ayz - a^3 bc - b^2 c - a c^2 =0 \,.
\end{equation}
The equations \eqref{L1}--\eqref{L2}, and consequently
\eqref{cr1}--\eqref{cr3}, make sense  if
\begin{equation}\label{abc}
 a > 0,~~~b > 0,~~~c > 0 \, .
\end{equation}

Inserting $\dot{x} = 0 = \dot{y} = \dot{z} =
\dot{a}=\dot{b}=\dot{c}$ into the left-hand-sides of
\eqref{cr1}--\eqref{cr6} leads to the set of equations
\begin{align}
  0 &= x^2 /a + b - a^3 \,, \label{c1}  \\
  0 &= y^2 /b + a^2 b - b^2 /a + c \,, \label{c2}  \\
  0 &= z^2 /c + a^2 c - c^2 /b \,, \label{c3} \\
  0 &= x \,, \label{c4} \\
  0 &= y \,, \label{c5} \\
  0 &= z \,. \label{c6}
\end{align}

For $x=0=y=z$ and $t < \infty$ the solution to
\eqref{c1}--\eqref{c3} does not exist if we insist on  \eqref{abc}
to be satisfied. Quite separate treatment is required for the case
$t \rightarrow \infty$:

Applying the substitution $t = 1/\tau$ to the system
\eqref{L1pol}--\eqref{L2pol} enables examination of that dynamics
in the limit $~\tau \rightarrow 0~$ instead of $t \rightarrow
\infty$, in the context  of the gravitational singularity.
However, that substitution turns \eqref{L1pol}--\eqref{L2pol} into
a system explicitly dependent on $\tau$. The latter prevents
making use of the dynamical systems analysis. Nevertheless,
another method is available.

First, it directly follows from \eqref{c1}--\eqref{c6}
that when the universe collapses, which includes $a\to 0^+$, then
$b/a\to 0^+$ and $c/b\to 0^+$, whence  we have $a\gg b\gg c >0$
in the neighbourhood of the singularity. A more precise estimation
can be obtained by introducing $\varepsilon_1
= b/a$ and $\varepsilon_2 = c/b$. This substitution turns
\eqref{c1}--\eqref{c3}, for $x=0=y=z$, into the system
\begin{align}
  0 &= a(\varepsilon_1-a^2) \,, \label{cc1}  \\
  0 &= a^3 (\varepsilon_1 - \varepsilon_1^2) + a\;\varepsilon_1\varepsilon_2 \,, \label{cc2}  \\
  0 &= a^3 \varepsilon_1\varepsilon_2 - a\; \varepsilon_1 \varepsilon_2^2 \,, \label{cc3}
\end{align}
which implies that $\varepsilon_1 = \varepsilon_2 =: \varepsilon$ and $a\sim\varepsilon^{1/2}$,
so that $b\sim\varepsilon^{3/2}$ and $c\sim\varepsilon^{5/2}$.  The latter is consistent with
the constraint \eqref{cr7}.  Our solution \eqref{solution} is obviously consistent with
this estimation.

Finally, the space of the critical points, $S$, turns out to be
\begin{equation}\label{crit}
  S =\{ (a,b,c,x,y,z)\in \dR^6~|~(x=0=y=z) \wedge(b=a^3) \} \, .
\end{equation}
Now, let us examine the type of the criticality of the elements of
the space $S$. Following the method of \cite{Per,Wig}, we first
determine the Jacobian, $J$, corresponding to the rhs of
\eqref{cr1}--\eqref{cr6}. It can be found to be
\begin{equation}\label{crJ}
 J=\begin{pmatrix}
 J_{11} & 1 & 0& 2x/a& 0 & 0  \\
 J_{21} & J_{22} & 1& 0& 2y/b & 0 \\
 2ac & c^2/b^2 & J_{33}& 0& 0 & 2zc \\
 0 & 0 & 0& 1&  0 & 0 \\
 0 & 0 & 0& 0&  1 & 0  \\
 0 & 0 & 0& 0& 0 & 1
 \end{pmatrix} \, ,
\end{equation}
where\\
$J_{11} = -x^2/a^2 - 3a^2,~~J_{21}= 2ab + b^2 / a^2,~~J_{22}= -
y^2/b^2 + a^2 - 2b/a ,$ and where $J_{33} = -z^2/c^2 + a^2 -2c/b
$.

The Jacobian evaluated at any point of $S$ and of order
$\varepsilon$  turns out to be

\begin{equation}\label{crJS}
 J_S =\begin{pmatrix}
 -3a^2 & 1 & 0& 0& 0 & 0 \\
 0 & -a^2 & 1& 0& 0 & 0 \\
 0 & 0 & -a^2& 0& 0 & 0 \\
 0 & 0 & 0& 1& 0 & 0 \\
 0 & 0 & 0& 0& 1 & 0 \\
 0 & 0 & 0& 0& 0 & 1
 \end{pmatrix} \, .
\end{equation}
Thus, the characteristic polynomial, $P(\lambda)$, associated with
$J_S$ reads
\begin{equation}\label{polyn}
  P(\lambda) = (-3a^2-\lambda)(-a^2-\lambda)^2(1- \lambda)^3 \, .
\end{equation}
It is clear that all the eigenvalues of $J_S$ are real numbers, as
$a \in \dR$, so that the space $S$ consists of the hyperbolic
critical points. However, in the limit $t \rightarrow \infty$ the
three eigenvalues vanish. Therefore, in the latter case the points
of $S$ are nonhyperbolic\footnote{A critical point is called a
hyperbolic critical point if all eigenvalues of the Jacobian
matrix of the linearised equations at this point have nonzero real
parts. Otherwise, it is called a nonhyperbolic critical point
\cite{Per,Wig}.}.

The dynamical systems analysis shows that the exact form of the dynamics
\eqref{L1}--\eqref{L2} includes, near the singularity, the instabilities connected
with the space of the nonhyperbolic critical points.
The dynamics near these points bifurcates \cite{Per,Wig}.

\section{Conclusions}

The best prototype for the BKL
scenario was derived by Belinski, Khalatnikov, and Ryan \cite{bkl}
in the context of the BKL conjecture. It is defined by Eqs.
\!\eqref{L1}--\eqref{L2} and the initial data satisfying condition
$a \gg b \gg c > 0$ (see \cite{Belinski:2014kba,book} for
more details).

In this paper we present an analytical solution to that prototype.
It is the only regular solution in the sense of Painlev\'e.
The solution was, to our best knowledge, unknown before. The evolution
presented in two recent papers \cite{Nick1,Nick2} are just numerical
simulations of the related dynamics in the Bianchi IX model case.

Our special monotonic solution \eqref{solution} is generically
unstable against perturbations of the initial data
\eqref{initial}. Interpreting the evolution of our system in two
interesting time directions, we see that (i) in the evolution
towards the singularity, the oscillating correction becomes
dominant compared to the monotonically decreasing solution
\eqref{solution}, and  (ii) in the direction away from the
singularity, the special solution \eqref{solution} and the uniform
expansion proportional to $K_3$ play the dominant role. In both
cases $K_3$ has to be sufficiently large in comparison with $K_1$
and $K_2$ to ensure monotonicity of the volume $V$.

These are perturbations of the solution to the dynamics derived
from the asymptotic dynamics of the {\it nondiagonal} Bianchi VIII
and Bianchi IX models, which underly the BKL scenario. Our
two-region scenario is similar to, but much more general than the
two stage evolution of the {\it diagonal} Bianchi IX model
considered by Grishchuk et al. \cite{LPG}. It is based on the
decomposition of the Bianchi IX metric into  the Friedmann
background (among other results) and the terms representing
gravitational waves.

 Exact solitonic gravitational perturbations on the Friedmann
background have been recently considered by Belinski et al. (see
\cite{BV} and references therein). They have shown that the
solitonic perturbations decay into gravitational waves during the
evolution away from the singularity. An interpretation of the
perturbations \eqref{asymp} as possible seeds of gravitational
waves will be published elsewhere \cite{GP}.

Making use of the dynamical systems techniques, we show that the
evolution of the system \eqref{L1}--\eqref{L2} is regular (can be
locally linearised) for any finite value of the evolution
parameter $t$. The dynamics approaches the space of nonhyperbolic
critical points in the limit $t \rightarrow \infty$, which means
that locally the dynamics cannot be linearised near those points.
In that limit the directional scale factors go to zero so that the
space volume $V$ goes to zero as well. The latter means (see
\cite{Nick1, Nick2} for more details) that the system approaches
the gravitational singularity.

It results from sections II and III that, as the system approaches
the singularity, its dynamics becomes sensitive to the choice of
the initial data, which means that the dynamics becomes chaotic.
We have shown that there exists a family of regular solutions to
the BKL system, which are parameterised by the real number $t_0
\in \dR $.  These solutions decay into chaos due to their
instabilities.  Thus,  chaotic behaviour is an inevitable companion
of the approach to the cosmological singularity. The latter is
consistent with the BKL scenario.

The existence of generic singular behaviour, predicted by BKL,
means that general relativity is not a complete theory of
gravitation. It is commonly expected that quantum gravity (still
to be constructed) would be free from the singularities. The
quantization of the BKL scenario carried out recently
\cite{AWG,AW} gives support to such expectation. However, this
result should be confirmed by quantizing that scenario, possibly
within a completely different approach, to verify its robustness.
As commonly known, quantization of a gravitational system is an
ambiguous procedure.

\acknowledgments We would like to thank Vladimir Belinski for
the valuable discussions.

\appendix

\section{Derivation of the solution (\ref{solution})}

To derive the solution, we are going to use singularity analysis.
Similar work was performed for the vacuum Bianchi IX matter
(mixmaster universe) in \cite{LCM}; one of their results was
regaining Taub's special solutions \cite{Taub1}.

The analysed equations  \eqref{L1} in their polynomial form read
\bs
\bl\label{L1pol} &a\, \ddot{a} - \dot{a}^2-a\, b+a^4=0 \, ,\\
&a\, b\, \ddot{b} - a\, \dot{b}^2 -a^3 b^2 + b^3 - a\, b\,
c=0 \, ,\\&b\, c\, \ddot{c}- b\, \dot{c}^2 - a^2 b\, c^2 + c^3 = 0,
\label{L1polc}
\end{align} while
the constraint \eqref{L2}  takes the polynomial form
\bl\label{L2pol} c\, \dot{a}\,\dot{b} + b\, \dot{a} \dot{c} + a\, \dot{b}\,\dot{c}-a^3b\, c -b^2
c-a\, c^2=0 \, .\end{align}
\es

The constant coefficients of the system
(\ref{L1pol})--(\ref{L2pol}) are obviously free of singular
points. Hence the positions of the possible singularities of the
solution are determined by the initial conditions (``movable
singularities'' \cite{CM,Conte}). To find a solution with a
singularity, which is a pole and thus does not introduce branching
{(i.e. has the Painlev\'e property), we proceed in a similar way
to that used in the classical Painlev\'e test \cite{ARS1,ARS2}.
Since our system is overdetermined, we need a more involved
analysis.

First, we look for a solution in the form of the Laurent series
about the assumed pole $t_0$ so that we have
\bs\label{Laur}
\bl
&a (t) =(t-t_0)^p\sum\limits_{n=0}^\infty a_n(t-t_0)^n,\\
&b (t) =(t-t_0)^q\sum\limits_{n=0}^\infty b_n(t-t_0)^n,\\
&c (t) =(t-t_0)^s\sum\limits_{n=0}^\infty c_n(t-t_0)^n.
\end{align}
\es

Substituting series \eqref{Laur} to the system
(\ref{L1pol})--(\ref{L2pol}), we obtain, in the zero order,
conditions of balance of the dominant terms. These conditions may
be satisfied in all equations (\ref{L1pol})--(\ref{L2pol})
provided that the exponents are $p=-1,~{q=-3},~s=-5$ (the other
family of solutions with $p=\!-1,\,{q=\!2},\,{s=\!0}$, does not
comply with the physics of
the system, as some of the terms are complex for real time).

A search for the initial exponents (only) in the Laurent expansion
was performed in \cite{SzydBies}. However, to our best knowledge,
the Painlev\'e analysis for the BKL equations
(\ref{L1pol})--(\ref{L2pol}) or other models including matter, has
never been done.

The coefficients at the dominant terms, obtained from the balance
equations prove to be
\be\label{sol0}
a_0=3,\quad b_0=30,\quad c_0=120.
\ee
In the Painlev\'e test, the higher order coefficients are obtained
from algebraic linear recurrence relations, which yield the
coefficients $a_n,\,b_n,\,c_n$ as functions of the lower order
coefficients and $t_0$. If the general solution was of the type
\eqref{Laur}, then the number of arbitrary constants should be
five, as the system consists of three \iind order ordinary
differential equations (ODE), \eqref{L1pol}--\eqref{L1polc}, with
one constraint \eqref{L2pol}. Bearing in mind that one of the
arbitrary constants is $t_0$, we see that the recurrence equations
should provide 4 more constants.

The arbitrary constants occur at the recurrence relation
determining those terms of the series \eqref{Laur} if two
conditions are satisfied together: (i) the rank of the coefficient
matrix is lower than the number of unknown coefficients and (ii)
the system is compatible, i.e. extending the coefficient matrix
with the right-hand sides of the algebraic equations does not
increase its rank.

In the Painlev\'e test, the indices of these terms are referred to
as ``resonances'' \cite{ARS1} or simply ``indices'' \cite{CM} (the
latter by analogy with the Fuchsian theory of linear differential
equations with singularities). For the first three equations
(\ref{L1pol})--(\ref{L1polc}) there is one positive ``resonant
index'' (a compromise between these two names) ${r_1=2}$, such
that the determinant of the algebraic recurrence system of 3
equations vanishes at $n=r_1=2$. However if we also require
satisfaction of the constraint \eqref{L2pol}, then the rank of the
coefficient matrix of the linear system is always 3, i.e. there
are $3\times 3$ submatrices of the coefficient matrix in
(\ref{L1pol})--(\ref{L2pol}) whose determinants are nonzero. The
requirement that the $3\times 3$ determinants vanish, has more
solutions: a resonant index at $r_2=-1$, which corresponds to the
arbitrariness of $t_0$, and four complex values:
\bs\label{indices}
\bl
&r_{3,4}=\half \left(1\pm i\sqrt{95-24\sqrt{6}}\right)\text{ and }\\
&r_{5,6}=\half \left(1\pm
i\sqrt{95+24\sqrt{6}}\right)\label{ind56}.
\end{align}
\es
Due to the fact that we get these complex ``indices'' instead of
the actual positive integer resonant indices, the recurrence
relations do not generate new arbitrary constants. Hence,
a possible solution with the pole may only be a special one. The
general solution does not have the form of a Laurent series (the
equations are non-Painlev\'e, i.e. contain branch points or
essential singularities which introduce branching \cite{Conte}).
The first 3 equations are compatible with the \ivth one. Hence,
the system of the recurrence relations has exactly one solution at
each order.

It is easy to recognise solution \eqref{solution} in the
series \eqref{Laur} limited to the zero-order terms. Indeed, Eq.
\eqref{solution}} is a solution of the system \eqref{L1} and
satisfies the constraint \eqref{L2} both for $t>t_0$ and
${t<t_0}$. At the same time, it consists of the zero-order
coefficients \eqref{sol0} divided by the appropriate powers of
$t-t_0$. This result is compatible with the recurrence relations:
they yield ${a_n=0},\,b_n=0,$ and $c_n=0,$ for all $n>0$ (again,
provided that we require satisfaction of the constraint
\eqref{L2}).

A similar expansion can be performed in the neighbourhood of
$t=\infty$. It results in the conclusion that there are no
solutions which tend to infinity while $t\to\infty$. The dominant
terms for $a,~b,~c$ prove to be $\pm 3t^{-1},~\pm 30t^{-3}$ and
$\pm 120t^{-5}$, respectively, corresponding to those of the
special solution \eqref{solution}. These are the only possible
coefficients if $t=\infty$ is a regular point or a pole. There are
no indices which are resonant for the whole 4-equation system,
i.e. the rank of the $4\times 3$ matrix of the coefficients is
never less than 3. The equations following from the requirement
that the determinants of all $3\times 3$ submatrices of the
coefficients vanish, yield solutions which cannot be the indices
of the expansion, namely $r=-1$ and the same irrational complex
$r$ as in \eqref{indices}. This way, the large-$t$ expansion
provides the same special solution \eqref{solution} as the
expansion about a hypothetic movable pole $t_0$.

An expansion in an arbitrary function $\Phi(t)$ instead of
$t-t_0$, shows that the special solution \eqref{solution} is the
only solution of \eqref{L1} which has the Painlev\'e property and
satisfies constraint \eqref{L2}. Another family of solutions of
\eqref{L1} having the Painlev\'e property exists, which are
proportional to powers of $\mathrm{cosec}\big(C(t-t_0)\big)$,
instead of $(t-t_0)^{-1}$, but they are incompatible with
constraint \eqref{L2}. The situation is similar to that obtained
in \cite{LCM},
although the model is different.

The Painlev\'e property is a usual companion to integrability and
regular behaviour, which includes lack of bifurcations (or
multifurcations) at unknown moments, but the connection is not
1:1. However this somewhat vague statement may be made stronger in
our case, because our system is autonomous and its physically
relevant solutions depend on $t$ through $t-t_0$ with
$t\in\mathbb{R}$. Hence also $t_0\in\mathbb{R}$ may be poles for
this class of solutions. This means that a physical non-Painlev\'e
solution \textit{has to} encounter the branching singularity in
its evolution towards $t_0$ or stem from the singularity in its
evolution from $t_0$.

\section{Monotonicity of volume}

This Appendix contains discussion of monotonicity of the volume.
We first discuss how the volume is affected by the linear
evolution of a small perturbation $\al,\,\beta,\,\g$ of
(respectively) $a,\, b,\, c$ given by solution \eqref{solution}.
Then, we add a short discussion of the general case. In the first
part we bear in mind that the linear approximation is valid in a
limited interval of time as the system is linearly unstable.

The volume $V=abc$ given by \eqref{Vol} should be a decreasing
function of time $t$ or $\th=\ln(t-t_0),~~ t>t_0$ (which also
means increasing in the backward evolution). In the linear
approximation, the $\th$-derivative of $V$ may be written as
\bl
&V'(\th)=-2160 e^{-17\th/2}\Bigg[150 K_3 e^{-3\th/2}+45 e^{-\th/2}
\nn\\
&+\sqrt{3}\left(\sqrt{152+53\sqrt{6}}\,
K_1\cos(\om_1\th+\psi_1)+\sqrt{152-53\sqrt{6}}\,
K_2\cos(\om_2\th+\psi_2)\right)\Bigg]\label{Vm},
\end{align}
where $\psi_1$ and $\psi_2$ are phases, which may be expressed in
terms of the constants $K_1,\,K_2,\,\ph_1,\,\ph_2$ (through the
classical replacements of expressions like $a\cos\al + b\sin\al$
by $\sqrt{a^2\!+b^2}\cos\big[\al \pm\arctan\,(b/a)\big]$ ).

The infimum of \eqref{Vm} $V'_{inf}$ corresponds to the two
cosines equal to $\sign(K_1)$ and $\sign(K_2)$ respectively, its
supremum $V'_{sup}$ to the respective $-\sign(K_1)$ and
$-\sign(K_2)$. Due to incommensurability of $\om_1$ and $\om_2$,
$V'$ may get arbitrarily close to its infimum or supremum if we
wait sufficiently long. The sufficient condition for monotonicity,
i.e. $V'<0$ for all $\th$, requires that (for all possible $\th$)
both $V'_{inf}<0$ and $V'_{sup}<0$. The latter is greater, hence
it is sufficient to examine $V'_{sup}$, as more demanding. It may
be written in the variable $\xi=e^{-\th/2}>0$ as
\bl\label{Vtrinom}
V'_{sup}&=-2160 \xi^{17}\left(150K_3 \xi^3+45\xi - r\right), \text{ where }\nn\\
 r~&=\sqrt{3}\left(\sqrt{152+53\sqrt{6}}\,\lvert
K_1\rvert+\sqrt{152-53\sqrt{6}}\,\lvert K_2\rvert\right)>0.
\end{align}
Real values of $\th$ correspond to $\xi>0$, where $\xi\to 0^+$
corresponds to $\th\to\infty$, while $\xi\to\infty$ corresponds to
$\th\to -\infty$. For $\xi>0$, the sign of $V'_{sup}$ follows from
the well known properties of third-degree algebraic equations.
Namely
\bi\item
For $K_3=0$ (non-generic) $V'_{sup}<0$ iff $\xi>r/45$, i.e.
$\th<-2\ln(r/45)$.
\item
For $K_3>0,~~V'_{sup}$ has one real zero $\xi_1>0$ (given by the
Cardano formula). As in the previous case, $V'<0$ iff $\xi>\xi_1$,
i.e. $\th<-2\ln(\xi_1)$.
\item
For $K_3<0, ~~V'_{sup}$ may have one positive real zero (for
$K_3\le -900/r^2$) or two positive real zeros (for
$-900/r^2<K_3<0$). However in this case $V'_{sup}<0$ requires that
$\th$ be \textit{above} some value, which affects the initial
stage of the evolution with non-physical oscillations of the
volume.
\ei
To summarise, there is no range of the considered constants which
ensure negative sign of $V'$ for all $\th$. However, if $K_3>0$, a
right-bounded interval of time (or $\th$) exists, in which $V'<0$.
This is sufficient for ensuring non-oscillatory behaviour of the
volume, provided that the right endpoint of this interval lies
beyond the range of the linear approximation.

 If we consider the evolution of $V$ without making the linear
approximation, then the summation of the BKL equations yields
\be
\frac{\dot{V}(t)}{V(t)} = \frac{d\ln\,(abc)}{dt}=
\frac{\dot{V}(t_{ini})}{V(t_{ini})}+\int_{t_{ini}}^t\,dt'\,a^2.
\ee
If the time-derivative $\dot{V}(t_{ini})>0$ at the initial time
$t_{ini}$, then the volume is an increasing function of $t$ (or
$\th$) for all $t>t_{ini}$. On the other hand, if we start from a
state of decreasing $V$, i.e. $\dot{V}(t_{ini}) <0$, then $V$
would further decrease towards some minimum. The question whether
this minimum is ever attained may be easily answered negatively by
looking at the diagonalised ``kinetic energy'' \eqref{cone1}. It
follows from the orthogonalization transformation \eqref{ortho}
that its first component is directly related to the volume $V$,
namely
\be
u_1=\tfrac13(x_1+x_2+x_3)=\tfrac13 \ln(V).
\ee
According to the inequality \eqref{cone1}, the ``kinetic energy''
has to be positive and may turn to zero at the singularity only.
This requires $\dot{u}_1^2>0$ everywhere outside the singularity,
i.e. initially negative $\dot{u}_1$ has to remain negative down to
the singularity. Hence the initially decreasing volume remains
decreasing throughout the whole evolution, q.e.d.

Note that this property is a consequence of constraint \eqref{L2}.

\end{document}